\documentclass[prl,twocolumn,amsmath,floats,showpacs,superscriptaddress]{revtex4}

\usepackage{graphicx}
\usepackage{dcolumn}

\begin{document}

\def\g0{$G_{0}$}

\title{Atomic size oscillations in conductance histograms for Au nanowires and the influence of
work hardening}

\author{I.K. Yanson}
\affiliation{B. Verkin Institute for Low Temperature Physics \&
Engineering, 47 Lenin Av., 61103, Kharkiv, Ukraine}
\affiliation{Kamerlingh Onnes Laboratorium, Universiteit Leiden,
Niels Bohrweg 2, 2333 CA Leiden, Netherlands}

\author{O.I. Shklyarevskii}
\affiliation{B. Verkin Institute for Low Temperature Physics \&
Engineering, 47 Lenin Av., 61103, Kharkiv, Ukraine}
\affiliation{Institute for Molecules and Materials, University of
Nijmegen, Toernooiveld 1, 6525 ED Nijmegen, Netherlands}

\author{Sz. Csonka}
\affiliation{Department of Physics, Institute of Physics, Budapest
University of Technology and Economics, 1111 Budapest, Hungary}

\author{H.~van~Kempen}
\affiliation{Institute for Molecules and Materials, University of
Nijmegen, Toernooiveld 1, 6525 ED Nijmegen, Netherlands}

\author{S. Speller}
\affiliation{Institute for Molecules and Materials, University of
Nijmegen, Toernooiveld 1, 6525 ED Nijmegen, Netherlands}

\author{A.I. Yanson}
\affiliation{Kamerlingh Onnes Laboratorium, Universiteit Leiden,
Niels Bohrweg 2, 2333 CA Leiden, Netherlands}

\author{J.M. van Ruitenbeek}
\affiliation{Kamerlingh Onnes Laboratorium, Universiteit Leiden,
Niels Bohrweg 2, 2333 CA Leiden, Netherlands}

\begin{abstract}
Nanowires of different nature have been shown to self-assemble as
a function of stress at the contact between two macroscopic
metallic leads. Here we demonstrate for Au wires that the balance
between various metastable nanowire configurations is influenced
by the microstructure of the starting materials and we discover a
new set of periodic structures, which we interpret as due to the
atomic discreteness of the contact size for the three principal
crystal orientations.
\end{abstract}

\date{\today}

\maketitle

Metallic nanowires have shown rich properties in terms of
self-organization phenomena that are controlled by a combination
of the quantum nature of the conduction electrons and the
atomic-scale surface energy (see review \cite{agrait03} and
references therein). Here we report yet another surprising series
of stable structures.

At the very smallest scale the metals Au, Pt and Ir, spontaneously
form into chains of atoms \cite{ohnishi98,yanson98,smit01}. For
slightly larger diameters further unusual arrangements referred to
as `weird wires' were predicted \cite{gulseren98} and later
observed in high-resolution transmission electron microscopy (TEM)
\cite{kondo00,oshima02,oshima03}. The observed structures for Au
have a helical arrangement in the form of concentric shells of
atoms.

In contrast to these atomic-packing driven structures, electronic
shell filling has been shown to lead to an independent series of
stable nanowire (NW) diameters for the free-electron-like alkali
metals \cite{yanson99,yanson00a}, and the noble metals
\cite{medina03,mares04,mares05}. These NWs were not imaged as in
TEM, but their stability was inferred from frequently occurring
stable conductance values during gentle breaking of the contacts
(see below). The series of stable values has a characteristic
period when plotted as a function of the square root of the
conductance, $\sqrt{G}$, which is a measure of the radius of the
wires.

Finally, regular bulk-packing of NWs has also been observed, where
the surface energy is the driving force, leading to completing of
flat facets of the wires. Such effects have been observed in TEM
\cite{kondo97,kizuka98,rodrigues00} as remarkably long and stable
wires, mostly for Au along the [110] direction. This atomic shell
filling series has also been observed in the conductance
\cite{yanson01a}, again as a regular period in $\sqrt{G}$.

Exactly which of these types of NWs self-assembles appears to
depend critically on the experimental conditions, which is not
fully understood. We anticipate that the selection of local minima
in the free energy is influenced by the dynamics of the wire
formation and the boundary conditions imposed by the structure of
the leads. Here, we present evidence that the microstructure of
the starting material, whether work hardened or annealed,
influences the appearance of a new series of stable NWs that are
periodic in the conductance, $G$, as opposed to $\sqrt{G}$ for the
NWs observed previously.

Gold is the archetypal metal when investigating quantum transport
phenomena in atomic-sized contacts \cite{agrait03}. The initial
interest into these systems was in quantization of the
conductance, in conjunction with the atomic discreteness of the
contacts. The experimental technique frequently employed, that we
also use here, involves the construction of histograms of
conductance values from a large number of contact-breaking traces.
The instrument used to adjust the contact size can be a scanning
tunnelling microscope or related piezo-controlled device. Here, we
employ the mechanically controllable break junction (MCBJ)
technique. Briefly, it involves breaking of a macroscopic wire of
the metal under study at low temperatures under cryogenic vacuum.
The clean fracture surfaces thus exposed are pushed back into
contact, and finely controlled indentation/retraction cycles can
be obtained through the action of a piezo-electric element. For a
more detailed description we refer to \cite{agrait03}.

The conduction histograms show the probability for observing a
given conductance value $G$. For Au one typically observes a
dominant peak in the histogram near $G= 1$ \g0, where
$G_{0}=2e^2/h$ is the conductance quantum. Further peaks are often
found near 2, 3, and 4--5 \g0,  both at cryogenic and at room
temperatures \cite{brandbyge95,ludoph99}. Slightly longer series
of peaks, with small peaks near $G=6$, and 7 \g0\ , have been
reported \cite{costa97a}, but these appear to be exceptional.
Although many authors have presented this series of peaks as
evidence for conductance quantization in Au, it has been pointed
out many times that one cannot make a straightforward separation
of genuine electronic quantization effects from discreteness due
to the atomic structure of the contacts \cite{agrait03,dreher05}.
It has now been generally accepted that the first peak results
from a one-atom contact, that also has a single conductance
channel. The higher peaks are predominantly due to stable
configurations of $n$ atoms in cross section, with some
quantization effects still observable \cite{ludoph99,brom99}.

\begin{figure}[!t]
\includegraphics[width=8cm,angle=0]{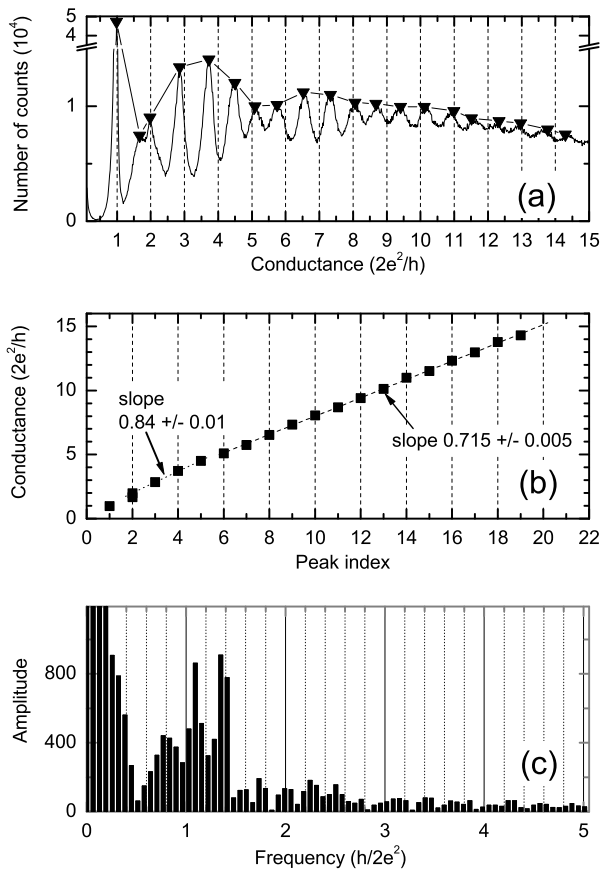}
\caption[]{Conductance histogram (a) obtained from 18165 curves
recorded while breaking the contact, using work hardened Au wires
at T = 6.5 K, and at a bias voltage of 80 mV. The connected
triangles show the position of oscillation maxima and emphasize
the beating pattern. The positions of the maxima in the histogram
are shown in (b) as a function of the index number, illustrating
the periodicity. Linear fits are shown by dashed lines. (c) The
components contributing to the beating pattern are obtained by a
Fourier transform of the data in (a) between 1.2 and 15 \g0 so as
to exclude the strong peak at 1 \g0. Three 'frequencies' can be
identified, $\sim$0.8; 1.1, and 1.4 $G_{0}^{-1}$.}\label{fig1}
\end{figure}

For the results reported here the starting wire materials are work
hardened Au wires, which is the key feature of our experiments.
The original wires were obtained commercially \cite{gold-wires},
but the results have been reproduced by pulling similar wires at
room temperature through a series of sapphire dies, reducing the
diameter in small steps from 500 $\mu$m to 100 $\mu$m, without
annealing the resulting wires. Using such wires we repeatedly
observe a periodic structure having up to 19 oscillations in the
conductance range from 0 to 15 \g0 (Fig.~\ref{fig1}(a)), with a
distinct superstructure modulation. Note that the axis is broken
and that the peak at 1 \g0 is about five times higher than the
other peaks. The low-conductance part of the histogram is similar
to those obtained in previous works on Au
\cite{brandbyge95,costa97a,ludoph99}, except that the peaks near 3
and 4 are much stronger, and larger than the one at 2 \g0. The
most striking feature is the large number of regularly spaced
oscillations, having a pronounced amplitude modulation.
Figure~\ref{fig1}(b) emphasizes that the oscillations are periodic
in $G$, not in $\sqrt{G}$. Figure~\ref{fig1}(c) shows a Fourier
transform of the spectrum in (a) between 1.2 and 15 \g0. We find
two prominent `frequencies', at $f_1 = 1.4$ and $f_2 = 1.1\,
G_{0}^{-1}$, that give rise to the beat minima near 2, 5, and 8
\g0. A smaller peak is visible at $f_3 = 0.8\,G_{0}^{-1}$.  In
order to see how the spectrum changes with thinning of the NWs we
performed a Fourier transform for different parts of original
curve. When the neck is thick (for $G=8-15\,G_{0}$), only the
$f_{1}$ periodicity is obtained. With further thinning the
admixture of frequency $f_{2}$ starts to emerge until its
amplitude approximately equals that of $f_{1}$. At the last stages
just before arriving at a single-atom contact a hint of the $f_{3}
$ periodicity appears.

\begin{figure}[!b]
\includegraphics[width=6cm,angle=0]{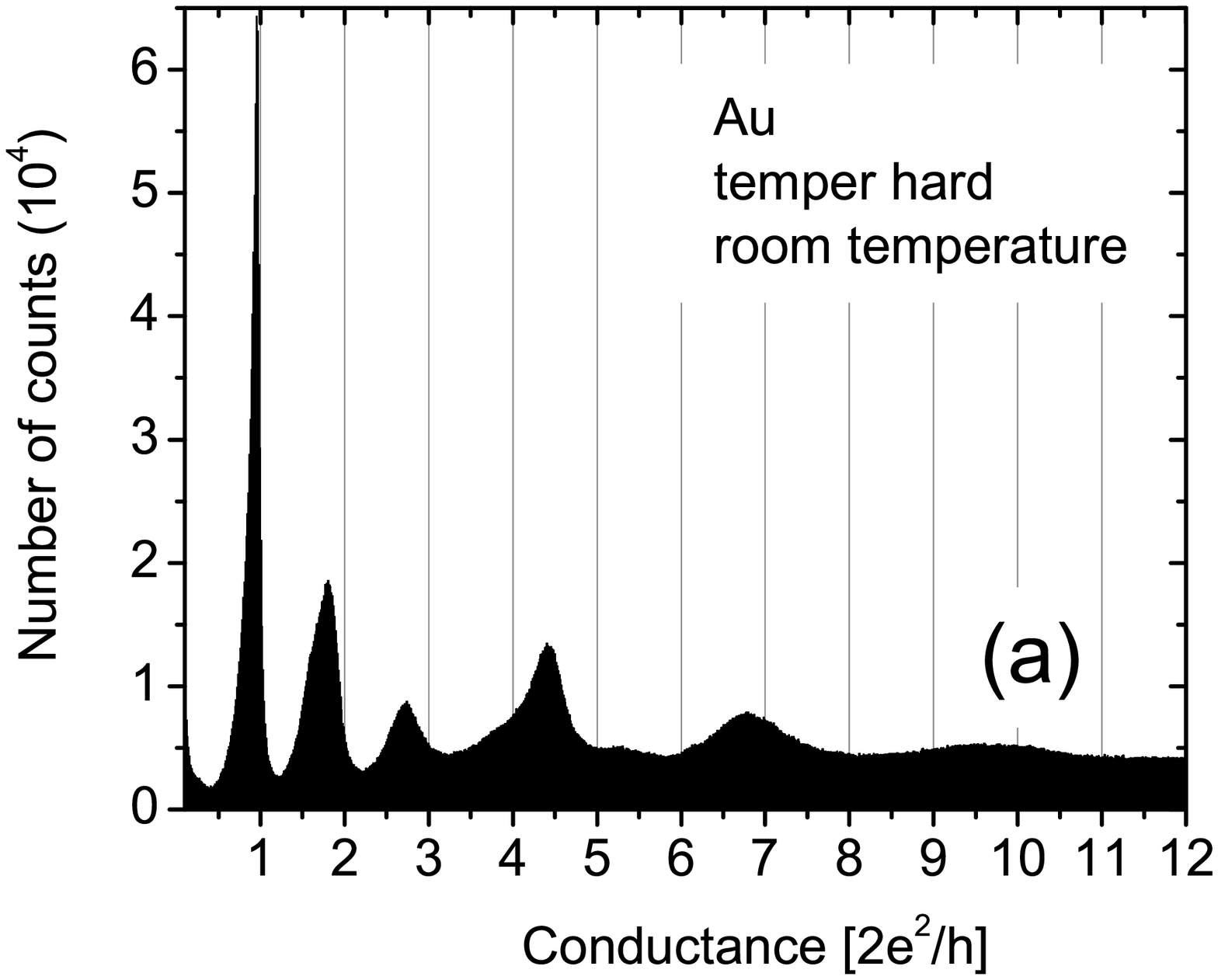}
\includegraphics[width=6cm,angle=0]{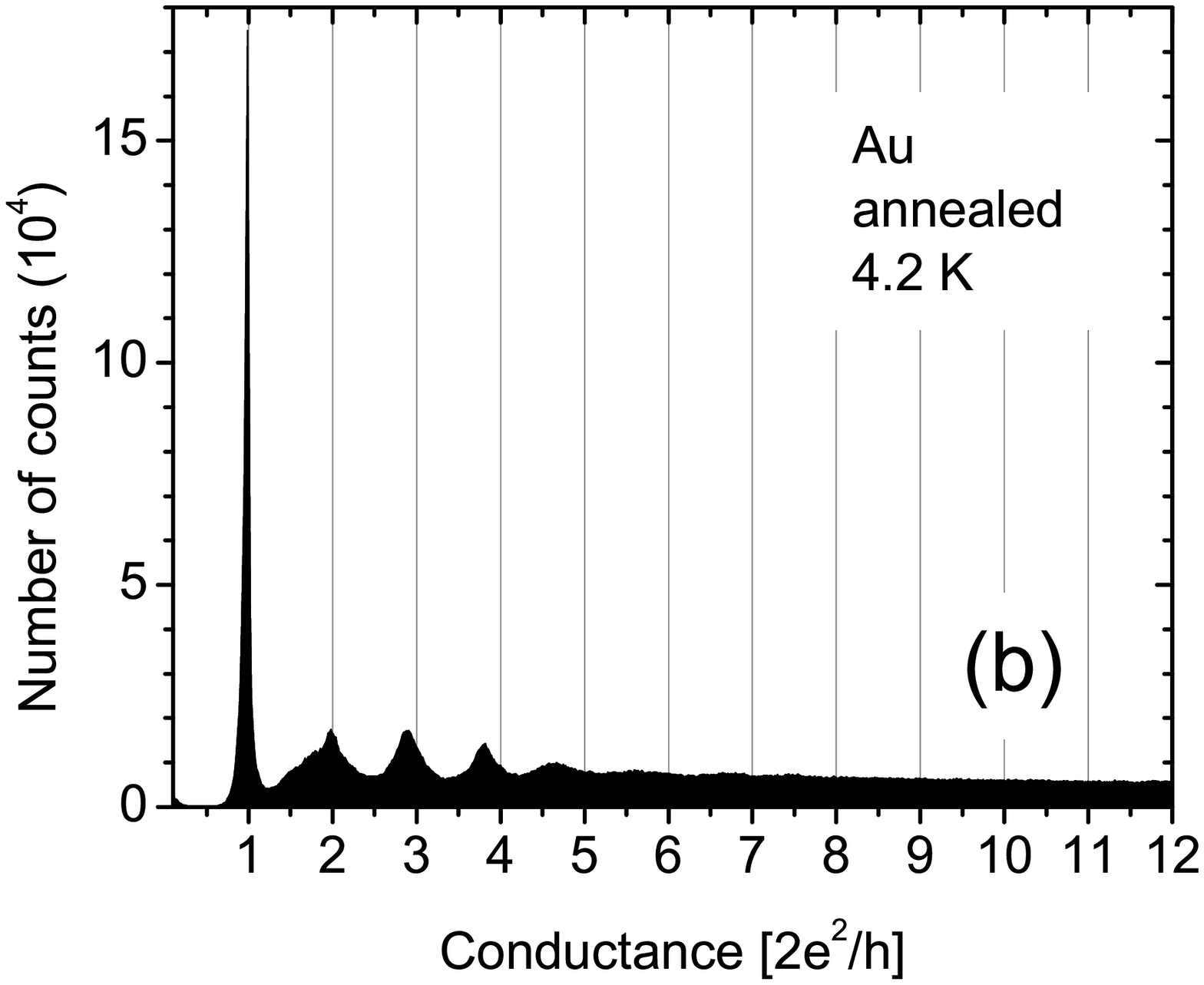}
\caption[]{Conductance histograms for work hardened Au wires (a)
at room temperature, including 38000 curves recorded at a bias
voltage of 80 mV. (b) Conductance histogram obtained at 4.2 K
after annealing the wires for 72 hours at 960 C. The number of
curves included is 15000 and the bias voltage was 80 mV.
}\label{fig2}
\end{figure}

Another important difference of the observed periodic structure in
Fig.~\ref{fig1}(a) with the atomic and electronic shell effects in
NWs is the fact that the present features are observed at
cryogenic temperatures, whereas shell effects are optimized by
raising the temperature to a significant fraction of the melting
point of the wires \cite{yanson01a,mares04,mares05}.
Figure~\ref{fig2}(a) shows a histogram for the same work hardened
Au wires recorded at room temperature. The histogram is very
similar to previously reported data for Au at room temperature
\cite{brandbyge95,costa97}. Some weak additional features are
discernable above 5 \g0 that are not periodic in $G$ and appear to
be mostly due to electronic shell structure, cf.\
\cite{mares04,mares05}. After annealing of the wires at elevated
temperatures the usual Au conductance histograms for cryogenic
temperatures \cite{costa97a,ludoph99} are recovered, as shown in
Fig.~\ref{fig2}(b). The amplitude modulation is absent and the
series of peaks terminates with very weak maxima at 6 or 7 \g0 .

The difference in mechanical properties of the contacts for
annealed and work hardened wires is also expressed in the global
variation of the conductance with stretching of the contact. In an
essay that measures the length of stretching between the point at
$G=15$ \g0 to the point of breaking over a large ensemble of
contact indentation/breaking cycles we observe that the mean
length is about 80\% {\em larger} for the work hardened wires. In
contrast, the relative height of first peak in the conductance
histogram is suppressed for work hardened wires (Fig.~\ref{fig1}
vs. Fig.~\ref{fig2}(b)), suggesting that chains of atoms are less
readily formed.

\begin{figure}[!b]
\includegraphics[width=6cm,angle=0]{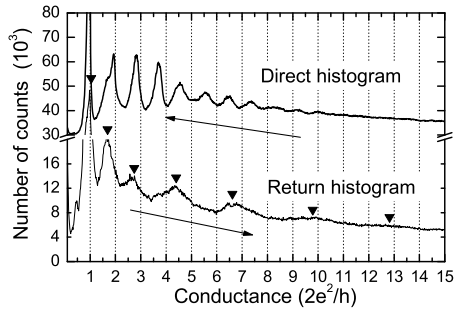}
\caption[]{Direct and return conductance histograms for a work
hardened Au sample at T=6.5 K. The direct histogram is shifted
vertically for clarity. Arrows show the direction of recording of
the conductance. Triangles show the positions of the maxima in the
return histogram.} \label{fig3}
\end{figure}

The properties observed depend on the direction of motion of the
electrodes. Fig.~\ref{fig3} shows the comparison of a direct and a
return conductance histogram for the same sample, which are
recorded while stretching or indenting the contact, respectively.
In the return histogram often (about 40\% of the histograms
recorded) $G$-periodic structure is replaced by a weak
$\sqrt{G}$-periodic structure reminiscent of the electronic shell
structure observed at room temperature \cite{mares04,mares05}.

Neither Cu nor Ag show a similar increase in the number of peaks
for work hardened wires.

We now turn to a discussion of these results. In our
interpretation we have been inspired by the results presented by
Rodrigues {\it et al.} \cite{rodrigues00}. Their room temperature
TEM images showed that Au NWs remain crystalline and fcc packed
during stretching of the contacts, down to the last stage of
thinning (see also Refs.~\cite{ohnishi98,kondo97,kizuka98}). The
authors argue that NWs tend to form along the three principal
crystallographic directions [100], [110], and [111]. Our key
finding is that the observed periods in the conductance histograms
may be explained by the atomic discreteness in the building of
wires along the principal directions.

The conductance of a NW can be expressed in terms of its
approximately circular cross-section area $A$ as
\cite{agrait03,torres94},
\begin{equation}
g=G/G_{0} = \pi A-(\pi A)^{1/2}+1/6 \label{area},
\end{equation}
which is valid in the semiclassical limit. Here $A$ is expressed
in units of the Fermi wavelength square, $\lambda _{F}^{2}$ and
the electrons are assumed to be confined by hard wall boundaries.
If we take spill out of about 0.34$\lambda _{F}$ beyond the
boundaries into account by a slightly larger diameter
\cite{ruitenbeek97}, the correction nearly cancels the last two
terms in Eq.~(\ref{area}). Therefore, we expect an approximately
linear relation between $g$ and $A$, $\Delta g \simeq \pi\, \Delta
A $. For fcc packing in layers perpendicular to the three
principal directions, [111], [100], and [110], one can identify
two-dimensional unit cells comprising two Au atoms each. The area
of the unit cells are and $\frac{1}{2} \sqrt{3}\,a^{2}$, $a^{2}$,
and $\sqrt{2}\,a^{2}$, respectively, where $a$ is the lattice
constant, $a=4.08$ \AA\ for bulk Au. Assuming the wire cross
section is incremented by a single atom the increment in
conductance for the NW would scale roughly as $\Delta g_{111} /
\Delta g_{100} / \Delta g_{110}  = \frac{1}{2} \sqrt{3} / 1 /
\sqrt{2}
 \cong 0.87/ 1 / 1.41 $. This is close to the ratio of the
periods, $1/f$, obtained from the experimental Fourier transform
in Fig.~\ref{fig1}(c). Quantitatively, along [100] the period
would be $\Delta g_{100}= \pi \,(\frac{1}{2}a^2/\lambda_{\rm F}^2)
= (k_{\rm F}a)^2/8\pi$. For a monovalent fcc metal $(k_{\rm
F}a)^3=12\pi^2$, giving $\Delta g_{100}= 0.96$, which is slightly
larger than the experimental period $(G_{0}/f_{2})=0.91$. For the
other two periods we obtain $\Delta g_{111}= 0.83$ and $\Delta
g_{110}= 1.36$, compared to experimental values of
$(G_{0}/f_{1})=0.71$ and $(G_{0}/f_{3})=1.25$. All experimental
values are slightly smaller than expected from this simplified,
clean free electron model. We tentatively attribute the reduction
of the periods to a small overall reduction of the conductance due
to scattering on defects near the contact, in analogy to similar
corrections in the observation of electronic and atomic shell
structure \cite{yanson99,yanson01a}. Further corrections may arise
from relaxation of the atoms at the surface with respect to the
fcc lattice positions.

We note that the relevant part of the NW at the smallest cross
section is likely to be short, probably of order of a lattice
spacing, $a$, in length. This may be deduced from the global shape
of the conductance curves as a function of stretching. For regular
wires at low temperatures the contact size decreases in steps
about a single atom wide until arriving at a single-atom contact.
The work hardened wires stretch almost twice this length, making
an average step width of about two lattice spacings. Yet, this is
long enough for Eq.~(\ref{area}) to apply.

One of the most unexpected aspects of our results is the influence
of work hardening of the starting materials, and the fact that
this enhances the structure in the conductance histograms at low
temperatures. The microstructure of the materials, in particular
the high density of grain boundaries may hold the answer to the
question why this occurs. As is know for nanocrystalline materials
(see, e.g. Ref.~\cite{wolf05}) atoms at grain boundaries tend to
flow under strain. This may provide enough freedom for the wires
to adjust the local orientation of the grains on either side of
the contact for optimal registry. At the same time there is not
enough thermal activation energy available at cryogenic
temperatures for atoms to diffuse over the surface and explore
other metastable NW configurations, such as those determined by
shell structure. The difference between stretching and compression
of the contacts (Fig.~\ref{fig3}) would follow in this
interpretation from the difference between the stable equilibrium
orientation of the two grains while pulling contrasted by the
unstable configuration of two grains touching at an apex while
being pushed into each other. For the latter we expect the grain
orientations to be pushed away from alignment and a grain boundary
to be introduced right at the contact point. The increased
mobility of atoms at this grain boundary may give rise to a weak
shell structure signal.

An extensive simulation of conductance histograms for Au
nanocontacts has been reported by Dreher {\it et al.}\
\cite{dreher05}. They indeed find evidence for atomic discreteness
in the histogram of contact cross sections, but this information
is smeared out in the conductance histogram. Although this point
deserves further study we speculate that the limited size of the
repeat cell in the simulation may lead to enhanced disorder in the
contact and a shorter NW length. This is partly visible in the
width of the first conductance peak, that is much larger than in
experiments.

\acknowledgments{This work is part of the research program of the
``Stichting FOM,'', partly sponsored through the SONS Programme of
the European Science Foundation, which is also funded by the
European Commission, Sixth Framework Programme, and was supported
by the "Nano" program of the National Academy of Science of
Ukraine under project \#10.05-H. O.I.S. wishes to acknowledge a
FOM visitor's grant. }

\bibliographystyle{apsrev}
\bibliography{Au-many-peaks}

\begin{thebibliography}{27}
\expandafter\ifx\csname natexlab\endcsname\relax\def\natexlab#1{#1}\fi
\expandafter\ifx\csname bibnamefont\endcsname\relax
  \def\bibnamefont#1{#1}\fi
\expandafter\ifx\csname bibfnamefont\endcsname\relax
  \def\bibfnamefont#1{#1}\fi
\expandafter\ifx\csname citenamefont\endcsname\relax
  \def\citenamefont#1{#1}\fi
\expandafter\ifx\csname url\endcsname\relax
  \def\url#1{\texttt{#1}}\fi
\expandafter\ifx\csname urlprefix\endcsname\relax\def\urlprefix{URL }\fi
\providecommand{\bibinfo}[2]{#2}
\providecommand{\eprint}[2][]{\url{#2}}

\bibitem[{\citenamefont{Agra{\"\i}t et~al.}(2003)\citenamefont{Agra{\"\i}t,
  {Levy Yeyati}, and van Ruitenbeek}}]{agrait03}
\bibinfo{author}{\bibfnamefont{N.}~\bibnamefont{Agra{\"\i}t}},
  \bibinfo{author}{\bibfnamefont{A.}~\bibnamefont{{Levy Yeyati}}},
  \bibnamefont{and} \bibinfo{author}{\bibfnamefont{J.~M.} \bibnamefont{van
  Ruitenbeek}}, \bibinfo{journal}{Phys. Rep.} \textbf{\bibinfo{volume}{377}},
  \bibinfo{pages}{81} (\bibinfo{year}{2003}).

\bibitem[{\citenamefont{Ohnishi et~al.}(1998)\citenamefont{Ohnishi, Kondo, and
  Takayanagi}}]{ohnishi98}
\bibinfo{author}{\bibfnamefont{H.}~\bibnamefont{Ohnishi}},
  \bibinfo{author}{\bibfnamefont{Y.}~\bibnamefont{Kondo}}, \bibnamefont{and}
  \bibinfo{author}{\bibfnamefont{K.}~\bibnamefont{Takayanagi}},
  \bibinfo{journal}{Nature} \textbf{\bibinfo{volume}{395}},
  \bibinfo{pages}{780} (\bibinfo{year}{1998}).

\bibitem[{\citenamefont{Yanson et~al.}(1998)\citenamefont{Yanson, {Rubio
  Bollinger}, van~den Brom, Agra{\"{\i}}t, and van Ruitenbeek}}]{yanson98}
\bibinfo{author}{\bibfnamefont{A.~I.} \bibnamefont{Yanson}},
  \bibinfo{author}{\bibfnamefont{G.}~\bibnamefont{{Rubio Bollinger}}},
  \bibinfo{author}{\bibfnamefont{H.~E.} \bibnamefont{van~den Brom}},
  \bibinfo{author}{\bibfnamefont{N.}~\bibnamefont{Agra{\"{\i}}t}},
  \bibnamefont{and} \bibinfo{author}{\bibfnamefont{J.~M.} \bibnamefont{van
  Ruitenbeek}}, \bibinfo{journal}{Nature} \textbf{\bibinfo{volume}{395}},
  \bibinfo{pages}{783} (\bibinfo{year}{1998}).

\bibitem[{\citenamefont{Smit et~al.}(2001)\citenamefont{Smit, Untiedt, Yanson,
  and van Ruitenbeek}}]{smit01}
\bibinfo{author}{\bibfnamefont{R.~H.~M.} \bibnamefont{Smit}},
  \bibinfo{author}{\bibfnamefont{C.}~\bibnamefont{Untiedt}},
  \bibinfo{author}{\bibfnamefont{A.~I.} \bibnamefont{Yanson}},
  \bibnamefont{and} \bibinfo{author}{\bibfnamefont{J.~M.} \bibnamefont{van
  Ruitenbeek}}, \bibinfo{journal}{Phys. Rev. Lett.}
  \textbf{\bibinfo{volume}{87}}, \bibinfo{pages}{266102}
  (\bibinfo{year}{2001}).

\bibitem[{\citenamefont{G{\"u}lseren et~al.}(1998)\citenamefont{G{\"u}lseren,
  Ercolessi, and Tosatti}}]{gulseren98}
\bibinfo{author}{\bibfnamefont{O.}~\bibnamefont{G{\"u}lseren}},
  \bibinfo{author}{\bibfnamefont{F.}~\bibnamefont{Ercolessi}},
  \bibnamefont{and} \bibinfo{author}{\bibfnamefont{E.}~\bibnamefont{Tosatti}},
  \bibinfo{journal}{Phys. Rev. Lett.} \textbf{\bibinfo{volume}{80}},
  \bibinfo{pages}{3775} (\bibinfo{year}{1998}).

\bibitem[{\citenamefont{Kondo and Takayanagi}(2000)}]{kondo00}
\bibinfo{author}{\bibfnamefont{Y.}~\bibnamefont{Kondo}} \bibnamefont{and}
  \bibinfo{author}{\bibfnamefont{K.}~\bibnamefont{Takayanagi}},
  \bibinfo{journal}{Science} \textbf{\bibinfo{volume}{289}},
  \bibinfo{pages}{606} (\bibinfo{year}{2000}).

\bibitem[{\citenamefont{Oshima et~al.}(2002)\citenamefont{Oshima, Koizumi,
  Mouri, Hirayama, Takayanagi, and Kondo}}]{oshima02}
\bibinfo{author}{\bibfnamefont{Y.}~\bibnamefont{Oshima}},
  \bibinfo{author}{\bibfnamefont{H.}~\bibnamefont{Koizumi}},
  \bibinfo{author}{\bibfnamefont{K.}~\bibnamefont{Mouri}},
  \bibinfo{author}{\bibfnamefont{H.}~\bibnamefont{Hirayama}},
  \bibinfo{author}{\bibfnamefont{K.}~\bibnamefont{Takayanagi}},
  \bibnamefont{and} \bibinfo{author}{\bibfnamefont{Y.}~\bibnamefont{Kondo}},
  \bibinfo{journal}{Phys. Rev. B} \textbf{\bibinfo{volume}{65}},
  \bibinfo{pages}{121401} (\bibinfo{year}{2002}).

\bibitem[{\citenamefont{Oshima et~al.}(2003)\citenamefont{Oshima, Onga, and
  Takayanagi}}]{oshima03}
\bibinfo{author}{\bibfnamefont{Y.}~\bibnamefont{Oshima}},
  \bibinfo{author}{\bibfnamefont{A.}~\bibnamefont{Onga}}, \bibnamefont{and}
  \bibinfo{author}{\bibfnamefont{K.}~\bibnamefont{Takayanagi}},
  \bibinfo{journal}{Phys. Rev. Lett.} \textbf{\bibinfo{volume}{91}},
  \bibinfo{pages}{205503} (\bibinfo{year}{2003}).

\bibitem[{\citenamefont{Yanson et~al.}(1999)\citenamefont{Yanson, Yanson, and
  van Ruitenbeek}}]{yanson99}
\bibinfo{author}{\bibfnamefont{A.~I.} \bibnamefont{Yanson}},
  \bibinfo{author}{\bibfnamefont{I.~K.} \bibnamefont{Yanson}},
  \bibnamefont{and} \bibinfo{author}{\bibfnamefont{J.~M.} \bibnamefont{van
  Ruitenbeek}}, \bibinfo{journal}{Nature} \textbf{\bibinfo{volume}{400}},
  \bibinfo{pages}{144} (\bibinfo{year}{1999}).

\bibitem[{\citenamefont{Yanson et~al.}(2000)\citenamefont{Yanson, Yanson, and
  van Ruitenbeek}}]{yanson00a}
\bibinfo{author}{\bibfnamefont{A.~I.} \bibnamefont{Yanson}},
  \bibinfo{author}{\bibfnamefont{I.~K.} \bibnamefont{Yanson}},
  \bibnamefont{and} \bibinfo{author}{\bibfnamefont{J.~M.} \bibnamefont{van
  Ruitenbeek}}, \bibinfo{journal}{Phys. Rev. Lett.}
  \textbf{\bibinfo{volume}{84}}, \bibinfo{pages}{5832} (\bibinfo{year}{2000}).

\bibitem[{\citenamefont{Medina et~al.}(2003)\citenamefont{Medina, D{\'\i}az,
  Le{\'o}n, Guerrero, Hasmy, Serena, and Costa-Kr{\"a}mer}}]{medina03}
\bibinfo{author}{\bibfnamefont{E.}~\bibnamefont{Medina}},
  \bibinfo{author}{\bibfnamefont{M.}~\bibnamefont{D{\'\i}az}},
  \bibinfo{author}{\bibfnamefont{N.}~\bibnamefont{Le{\'o}n}},
  \bibinfo{author}{\bibfnamefont{C.}~\bibnamefont{Guerrero}},
  \bibinfo{author}{\bibfnamefont{A.}~\bibnamefont{Hasmy}},
  \bibinfo{author}{\bibfnamefont{P.~A.} \bibnamefont{Serena}},
  \bibnamefont{and} \bibinfo{author}{\bibfnamefont{J.~L.}
  \bibnamefont{Costa-Kr{\"a}mer}}, \bibinfo{journal}{Phys. Rev. Lett.}
  \textbf{\bibinfo{volume}{91}}, \bibinfo{pages}{026802}
  (\bibinfo{year}{2003}).

\bibitem[{\citenamefont{Mares et~al.}(2004)\citenamefont{Mares, Otte,
  Soukiassian, Smit, and van Ruitenbeek}}]{mares04}
\bibinfo{author}{\bibfnamefont{A.~I.} \bibnamefont{Mares}},
  \bibinfo{author}{\bibfnamefont{A.~F.} \bibnamefont{Otte}},
  \bibinfo{author}{\bibfnamefont{L.~G.} \bibnamefont{Soukiassian}},
  \bibinfo{author}{\bibfnamefont{R.~H.~M.} \bibnamefont{Smit}},
  \bibnamefont{and} \bibinfo{author}{\bibfnamefont{J.~M.} \bibnamefont{van
  Ruitenbeek}}, \bibinfo{journal}{Phys. Rev. B} \textbf{\bibinfo{volume}{70}},
  \bibinfo{pages}{073401} (\bibinfo{year}{2004}).

\bibitem[{\citenamefont{Mares and van Ruitenbeek}(2005)}]{mares05}
\bibinfo{author}{\bibfnamefont{A.~I.} \bibnamefont{Mares}} \bibnamefont{and}
  \bibinfo{author}{\bibfnamefont{J.~M.} \bibnamefont{van Ruitenbeek}},
  \bibinfo{journal}{Phys. Rev. B}  (\bibinfo{year}{2005}), \bibinfo{note}{in
  print; cond-mat/0506728}.

\bibitem[{\citenamefont{Kondo and Takayanagi}(1997)}]{kondo97}
\bibinfo{author}{\bibfnamefont{Y.}~\bibnamefont{Kondo}} \bibnamefont{and}
  \bibinfo{author}{\bibfnamefont{K.}~\bibnamefont{Takayanagi}},
  \bibinfo{journal}{Phys. Rev. Lett.} \textbf{\bibinfo{volume}{79}},
  \bibinfo{pages}{3455} (\bibinfo{year}{1997}).

\bibitem[{\citenamefont{Kizuka}(1998)}]{kizuka98}
\bibinfo{author}{\bibfnamefont{T.}~\bibnamefont{Kizuka}},
  \bibinfo{journal}{Phys. Rev. Lett.} \textbf{\bibinfo{volume}{81}},
  \bibinfo{pages}{4448} (\bibinfo{year}{1998}).

\bibitem[{\citenamefont{Rodrigues et~al.}(2000)\citenamefont{Rodrigues, Fuhrer,
  and Ugarte}}]{rodrigues00}
\bibinfo{author}{\bibfnamefont{V.}~\bibnamefont{Rodrigues}},
  \bibinfo{author}{\bibfnamefont{T.}~\bibnamefont{Fuhrer}}, \bibnamefont{and}
  \bibinfo{author}{\bibfnamefont{D.}~\bibnamefont{Ugarte}},
  \bibinfo{journal}{Phys. Rev. Lett.} \textbf{\bibinfo{volume}{85}},
  \bibinfo{pages}{4124} (\bibinfo{year}{2000}).

\bibitem[{\citenamefont{Yanson et~al.}(2001)\citenamefont{Yanson, Yanson, and
  van Ruitenbeek}}]{yanson01a}
\bibinfo{author}{\bibfnamefont{A.~I.} \bibnamefont{Yanson}},
  \bibinfo{author}{\bibfnamefont{I.~K.} \bibnamefont{Yanson}},
  \bibnamefont{and} \bibinfo{author}{\bibfnamefont{J.~M.} \bibnamefont{van
  Ruitenbeek}}, \bibinfo{journal}{Phys. Rev. Lett.}
  \textbf{\bibinfo{volume}{87}}, \bibinfo{pages}{216805}
  (\bibinfo{year}{2001}).

\bibitem[{\citenamefont{Brandbyge et~al.}(1995)\citenamefont{Brandbyge,
  Schi{\o}tz, S{\o}rensen, Stoltze, Jacobsen, N{\o}rskov, Olesen,
  L{\ae}gsgaard, Stensgaard, and Besenbacher}}]{brandbyge95}
\bibinfo{author}{\bibfnamefont{M.}~\bibnamefont{Brandbyge}},
  \bibinfo{author}{\bibfnamefont{J.}~\bibnamefont{Schi{\o}tz}},
  \bibinfo{author}{\bibfnamefont{M.~R.} \bibnamefont{S{\o}rensen}},
  \bibinfo{author}{\bibfnamefont{P.}~\bibnamefont{Stoltze}},
  \bibinfo{author}{\bibfnamefont{K.~W.} \bibnamefont{Jacobsen}},
  \bibinfo{author}{\bibfnamefont{J.~K.} \bibnamefont{N{\o}rskov}},
  \bibinfo{author}{\bibfnamefont{L.}~\bibnamefont{Olesen}},
  \bibinfo{author}{\bibfnamefont{E.}~\bibnamefont{L{\ae}gsgaard}},
  \bibinfo{author}{\bibfnamefont{I.}~\bibnamefont{Stensgaard}},
  \bibnamefont{and}
  \bibinfo{author}{\bibfnamefont{F.}~\bibnamefont{Besenbacher}},
  \bibinfo{journal}{Phys. Rev. B} \textbf{\bibinfo{volume}{52}},
  \bibinfo{pages}{8499} (\bibinfo{year}{1995}).

\bibitem[{\citenamefont{Ludoph et~al.}(1999)\citenamefont{Ludoph, Devoret,
  Esteve, Urbina, and van Ruitenbeek}}]{ludoph99}
\bibinfo{author}{\bibfnamefont{B.}~\bibnamefont{Ludoph}},
  \bibinfo{author}{\bibfnamefont{M.~H.} \bibnamefont{Devoret}},
  \bibinfo{author}{\bibfnamefont{D.}~\bibnamefont{Esteve}},
  \bibinfo{author}{\bibfnamefont{C.}~\bibnamefont{Urbina}}, \bibnamefont{and}
  \bibinfo{author}{\bibfnamefont{J.~M.} \bibnamefont{van Ruitenbeek}},
  \bibinfo{journal}{Phys. Rev. Lett.} \textbf{\bibinfo{volume}{82}},
  \bibinfo{pages}{1530} (\bibinfo{year}{1999}).

\bibitem[{\citenamefont{Costa-Kr{\"a}mer
  et~al.}(1997)\citenamefont{Costa-Kr{\"a}mer, Garc{\'\i}a, and
  Olin}}]{costa97a}
\bibinfo{author}{\bibfnamefont{J.~L.} \bibnamefont{Costa-Kr{\"a}mer}},
  \bibinfo{author}{\bibfnamefont{N.}~\bibnamefont{Garc{\'\i}a}},
  \bibnamefont{and} \bibinfo{author}{\bibfnamefont{H.}~\bibnamefont{Olin}},
  \bibinfo{journal}{Phys. Rev. B} \textbf{\bibinfo{volume}{55}},
  \bibinfo{pages}{12910} (\bibinfo{year}{1997}).

\bibitem[{\citenamefont{Dreher et~al.}(2005)\citenamefont{Dreher, Pauly,
  Heurich, Cuevas, Scheer, and Nielaba}}]{dreher05}
\bibinfo{author}{\bibfnamefont{M.}~\bibnamefont{Dreher}},
  \bibinfo{author}{\bibfnamefont{F.}~\bibnamefont{Pauly}},
  \bibinfo{author}{\bibfnamefont{J.}~\bibnamefont{Heurich}},
  \bibinfo{author}{\bibfnamefont{J.~C.} \bibnamefont{Cuevas}},
  \bibinfo{author}{\bibfnamefont{E.}~\bibnamefont{Scheer}}, \bibnamefont{and}
  \bibinfo{author}{\bibfnamefont{P.}~\bibnamefont{Nielaba}},
  \bibinfo{journal}{Phys. Rev. B} \textbf{\bibinfo{volume}{72}},
  \bibinfo{pages}{075435} (\bibinfo{year}{2005}).

\bibitem[{\citenamefont{van~den Brom and van Ruitenbeek}(1999)}]{brom99}
\bibinfo{author}{\bibfnamefont{H.~E.} \bibnamefont{van~den Brom}}
  \bibnamefont{and} \bibinfo{author}{\bibfnamefont{J.~M.} \bibnamefont{van
  Ruitenbeek}}, \bibinfo{journal}{Phys. Rev. Lett.}
  \textbf{\bibinfo{volume}{82}}, \bibinfo{pages}{1526} (\bibinfo{year}{1999}).

\bibitem[{gol()}]{gold-wires}
\bibinfo{note}{Gold wire, 75 $\mu$m diameter, ``temper hard'', Advent Research
  Materials Ltd, Catalogue No 518609}.

\bibitem[{\citenamefont{Costa-Kr{\"a}mer}(1997)}]{costa97}
\bibinfo{author}{\bibfnamefont{J.~L.} \bibnamefont{Costa-Kr{\"a}mer}},
  \bibinfo{journal}{Phys. Rev. B} \textbf{\bibinfo{volume}{55}},
  \bibinfo{pages}{R4875} (\bibinfo{year}{1997}).

\bibitem[{\citenamefont{Torres et~al.}(1994)\citenamefont{Torres, Pascual, and
  S{\'a}enz}}]{torres94}
\bibinfo{author}{\bibfnamefont{J.~A.} \bibnamefont{Torres}},
  \bibinfo{author}{\bibfnamefont{J.~I.} \bibnamefont{Pascual}},
  \bibnamefont{and} \bibinfo{author}{\bibfnamefont{J.~J.}
  \bibnamefont{S{\'a}enz}}, \bibinfo{journal}{Phys. Rev. B}
  \textbf{\bibinfo{volume}{49}}, \bibinfo{pages}{16581} (\bibinfo{year}{1994}).

\bibitem[{\citenamefont{van Ruitenbeek et~al.}(1997)\citenamefont{van
  Ruitenbeek, Devoret, Esteve, and Urbina}}]{ruitenbeek97}
\bibinfo{author}{\bibfnamefont{J.~M.} \bibnamefont{van Ruitenbeek}},
  \bibinfo{author}{\bibfnamefont{M.~H.} \bibnamefont{Devoret}},
  \bibinfo{author}{\bibfnamefont{D.}~\bibnamefont{Esteve}}, \bibnamefont{and}
  \bibinfo{author}{\bibfnamefont{C.}~\bibnamefont{Urbina}},
  \bibinfo{journal}{Phys. Rev. B} \textbf{\bibinfo{volume}{56}},
  \bibinfo{pages}{12566} (\bibinfo{year}{1997}).

\bibitem[{\citenamefont{Wolf et~al.}(2005)\citenamefont{Wolf, Yamakov,
  Phillpot, Mukherjee, and Gleiter}}]{wolf05}
\bibinfo{author}{\bibfnamefont{D.}~\bibnamefont{Wolf}},
  \bibinfo{author}{\bibfnamefont{V.}~\bibnamefont{Yamakov}},
  \bibinfo{author}{\bibfnamefont{S.~R.} \bibnamefont{Phillpot}},
  \bibinfo{author}{\bibfnamefont{A.}~\bibnamefont{Mukherjee}},
  \bibnamefont{and} \bibinfo{author}{\bibfnamefont{H.}~\bibnamefont{Gleiter}},
  \bibinfo{journal}{Acta Materialia} \textbf{\bibinfo{volume}{53}},
  \bibinfo{pages}{1} (\bibinfo{year}{2005}).

\end{thebibliography}

\end{document}